\documentclass[preprint2]{aastex}
\usepackage{geometry}                
\usepackage{graphicx}
\usepackage{amssymb}
\usepackage{amsmath}
\usepackage{multirow}
\usepackage{rotating}
\usepackage{graphicx}
\usepackage{bm}
\usepackage{color}

\begin{document}

\title{Type Ia Supernova Hubble Residuals and Host-Galaxy Properties}

\author
{
    A.~G.~Kim,\altaffilmark{1}
    G.~Aldering,\altaffilmark{1}
    P.~Antilogus,\altaffilmark{2}
    C.~Aragon,\altaffilmark{1}
    S.~Bailey,\altaffilmark{1}
    C.~Baltay,\altaffilmark{3}
    S.~Bongard,\altaffilmark{2}
    C.~Buton,\altaffilmark{4}
    A.~Canto,\altaffilmark{2}
    F.~Cellier-Holzem,\altaffilmark{2}
    M.~Childress,\altaffilmark{5} 
    N.~Chotard,\altaffilmark{6}
    Y.~Copin,\altaffilmark{6}
    H.~K. Fakhouri,\altaffilmark{1,7}
    U.~Feindt,\altaffilmark{4}
    M.~Fleury,\altaffilmark{2}
    E.~Gangler,\altaffilmark{6}
    P.~Greskovic,\altaffilmark{4}
    J.~Guy,\altaffilmark{2} 
    M.~Kowalski,\altaffilmark{4}
    S.~Lombardo,\altaffilmark{4}
    J.~Nordin,\altaffilmark{1,8}
    P.~Nugent,\altaffilmark{9,10}
    R.~Pain,\altaffilmark{2}
    E.~Pecontal,\altaffilmark{11}
    R.~Pereira,\altaffilmark{6}
    S.~Perlmutter,\altaffilmark{1,7}
    D.~Rabinowitz,\altaffilmark{3}
    M.~Rigault,\altaffilmark{6}  
    K.~Runge,\altaffilmark{1}
    C.~Saunders,\altaffilmark{1}
    R.~Scalzo,\altaffilmark{5}
    G.~Smadja,\altaffilmark{6}
    C.~Tao,\altaffilmark{12,13}
    R.~C. Thomas,\altaffilmark{8}
    B.~A.~Weaver\altaffilmark{14}
}

\altaffiltext{1}
{
    Physics Division, Lawrence Berkeley National Laboratory, 
    1 Cyclotron Road, Berkeley, CA, 94720
}
\altaffiltext{2}
{
    Laboratoire de Physique Nucl\'eaire et des Hautes \'Energies,
    Universit\'e Pierre et Marie Curie Paris 6, Universit\'e Paris Diderot Paris 7, CNRS-IN2P3, 
    4 place Jussieu, 75252 Paris Cedex 05, France
}
\altaffiltext{3}
{
    Department of Physics, Yale University, 
    New Haven, CT, 06250-8121
}
\altaffiltext{4}
{
    Physikalisches Institut, Universit\"at Bonn,
    Nu\ss allee 12, 53115 Bonn, Germany
}

\altaffiltext{5}
{
    Research School of Astronomy and Astrophysics,
    Australian National University,
    Canberra, ACT 2611, Australia
}

\altaffiltext{6}
{
    Universit\'e de Lyon, F-69622, Lyon, France ; Universit\'e de Lyon 1, Villeurbanne ; 
    CNRS/IN2P3, Institut de Physique Nucl\'eaire de Lyon.
}

\altaffiltext{7}
{
    Department of Physics, University of California Berkeley,
    366 LeConte Hall MC 7300, Berkeley, CA, 94720-7300
}

\altaffiltext{8}
{
Space Sciences Laboratory, University of California Berkeley, 7 Gauss Way, Berkeley, CA 94720, USA
}

\altaffiltext{9}
{
    Computational Cosmology Center, Computational Research Division, Lawrence Berkeley National Laboratory, 
    1 Cyclotron Road MS 50B-4206, Berkeley, CA, 94720
}

\altaffiltext{10}
{
        Department of Astronomy, University of California Berkeley,
        B-20 Hearst Field Annex \# 3411, Berkeley, CA, 94720-3411
}

\altaffiltext{11}
{
    Centre de Recherche Astronomique de Lyon, Universit\'e Lyon 1,
    9 Avenue Charles Andr\'e, 69561 Saint Genis Laval Cedex, France
}

\altaffiltext{12}
{
    Centre de Physique des Particules de Marseille , Aix-Marseille Universit\'e , CNRS/IN2P3, 163, avenue de Luminy - Case 902 - 13288 Marseille Cedex 09, France
}
\altaffiltext{13}
{
    Tsinghua Center for Astrophysics, Tsinghua University, Beijing 100084, China 
}
\altaffiltext{14}
{
    Center for Cosmology and Particle Physics,
    New York University,
    4 Washington Place, New York, NY 10003, USA
}

\begin{abstract}
\cite{Kim} [K13]  introduced a new methodology for determining peak-brightness absolute magnitudes of type Ia supernovae from
multi-band light curves.  We examine the relation between
their parameterization of 
light curves and Hubble residuals, based on photometry synthesized from the
Nearby Supernova Factory spectrophotometric time series, with global host-galaxy properties.
The K13 Hubble residual step with host mass is
 $0.013\pm 0.031$ mag for a supernova subsample with data coverage corresponding to the K13 training;
at $\ll 1\sigma$, the step is not significant and lower than previous measurements.
Relaxing the data coverage requirement the Hubble residual step with host mass is $0.045\pm 0.026$ mag for the larger sample;
a calculation using the modes of the distributions, less sensitive to outliers, yields a step of  0.019 mag.
The analysis of this article uses K13 inferred luminosities,
as distinguished from previous works that use magnitude corrections as a function of SALT2 color and stretch parameters:
Steps at $>2\sigma $ significance are found in SALT2 Hubble residuals in samples split
by the values of their K13 $x(1)$ and $x(2)$ light-curve parameters.
$x(1)$ affects the light-curve width and color around peak (similar to the $\Delta m_{15}$ and stretch parameters),
and $x(2)$ affects  colors, the near-UV light-curve width, and the light-curve decline 20 to 30 days after peak brightness.
The novel light-curve analysis, increased parameter set, and magnitude corrections of K13 may be capturing features of SN~Ia diversity
arising from progenitor stellar evolution. 
\end{abstract}

\keywords{distance scale, supernovae: general}

\section{Introduction}
Type Ia supernovae (SNe~Ia) serve as distance indicators used to measure the expansion history of the Universe.
Although supernovae are not perfect standard candles, the peak absolute magnitude of an individual event can be inferred from
observed multi-band light curves and a redshift using trained empirical relations.
SN~Ia optical light curves have homogeneous time evolution, which allowed them to be described by a template
 \citep[see][and references therein]{1991A&AS...89..537L}.
The relationship between light-curve decline rates and their correlation with absolute magnitude was noted by \citet{1984SvA....28..658P}
and further developed by \citet{1993ApJ...413L.105P}, and was confirmed with the supernovae observed by the Calan/Tololo Survey
\citep{1996AJ....112.2391H,1996ApJ...473...88R}.   An observed-color parameter was added to the modeling of multi-band light curves
\citep{1998A&A...331..815T}.
Today there is a suite of models that parameterize supernova light-curve shapes and colors,
which are used to standardize absolute magnitudes to within a seemingly random $0.10$--$0.15$ mag dispersion
\citep{ 2007A&A...466...11G, 2007ApJ...659..122J, 2008ApJ...681..482C, 2011AJ....141...19B,  2011ApJ...731..120M}.

The host galaxy  conveys information about the supernova progenitor environment.
Although they do not describe an individual star, the host mass, specific star formation rate, and metallicity 
provide an expectation of the progenitor  initial conditions that can be related to peak absolute magnitude.
Dependence of light-curve parameters and Hubble residuals (inferred
magnitudes from light curves minus those expected 
from the cosmological distance-redshift relation, or Hubble law) on {global}\footnote{Global properties
are drawn from measurements over a large solid angle of the host galaxy.  This contrasts with ``local'' properties
that are drawn from a small solid angle around the supernova position \citep{Rigault}.}
host-galaxy properties has been sought.
\citet{1989PASP..101..588F} showed and \citet{1996AJ....112.2438H} confirmed that the light-curve shape parameter
is correlated with host-galaxy morphology.
\citet{2010ApJ...715..743K,2010MNRAS.406..782S,2010ApJ...722..566L,2011ApJ...740...92G, 2012ApJ...746...85S} find that Hubble residuals
depend on host mass. \citet{2011arXiv1101.4269K} find a similar dependence on metallicity
while \citet{2011ApJ...743..172D,2013ApJ...764..191H} find a dependence on both metallicity and specific star formation rate (sSFR).

\citet[][hereafter C13b]{2013ApJ...770..108C} perform such an analysis on
the supernovae of the Nearby Supernova Factory  \citep[SNfactory,][]{2002SPIE.4836...61A}.
Supernova distances are derived using linear magnitude corrections
based on light-curve shape and color parameters from
SALT2 \citep[v2.2.0;][]{ 2007A&A...466...11G,2010A&A...523A...7G} 
fits to SNfactory synthetic photometry, using the procedure described in \citet{2009A&A...500L..17B};
in this article these linearly-corrected distances are referred to as ``SALT2'' distances.
Host mass,
sSFR, and metallicity are  derived from photometric and spectroscopic observations of the
associated galaxies \citep[][hereafter C13a]{2013ApJ...770..107C}.
Their findings are consistent with previous studies; when
splitting the SN Ia sample by host mass, sSFR, and metallicity at $\log{(M_\star/M_\sun)} = 10.0$,
$\log{(\mbox{sSFR})} =-10.3$,
and  $12 + \log{(\mbox{O/H})} = 8.8$ respectively, they find that SNe Ia in high-mass (low-sSFR, high-metallicity) hosts
are on average $0.085 \pm 0.028$ mag ($0.050 \pm 0.029$ mag, $0.103 \pm 0.036$ mag) brighter than those in low-mass (high-sSFR, low-metallicity) hosts after brightness corrections based on the SALT2 light-curve shape and color
brightness corrections.

The Hubble residuals depend on the model  used to determine absolute magnitude.
Although there is the expectation that  the progenitor variability tracked by
host-galaxy parameters must also be directly manifest within the supernova signal itself,
it appears not to be captured by the light-curve models used and the associated standardization in the cited work.
The SDSS-II Supernova Survey, using
samples
divided by passive and star-forming hosts, finds Hubble residual biases between both SALT2- and MLCS2k2-determined
distances \citep{2010ApJ...722..566L}: indication that the bias from the two light-curve fitters share a 
common source.  The two parameters of one model are highly correlated
with the two parameters of the other \citep{2009ApJS..185...32K}, which brings to question whether
a third light-curve parameter associated with host properties is not being captured by SALT2 or MLCS2k2.
Although there are searches for such a third light-curve parameter associated with Hubble residual bias (e.g.\
\citet{2010ApJ...712..350H} who test whether heterogeneity in light-curve rise times can account for the SDSS-II result),  as of
yet no such parameter has  been found.

\citet[][hereafter K13]{Kim} expand the optical light-curve parameterization by characterizing
light curves through the probability distribution function of a Gaussian process for the regressed values
at phases $-10$ to 35 in one-day intervals relative to peak, rather than the parameters
of a best-fit model.
The relationship between the K13 light-curve parameters  and light-curve shapes can be seen in Figure~4 of K13, and are
described briefly here.  The effect of the $x(0)$ parameter on 
the light curve is relatively phase-independent and is increasingly stronger in bluer bands, very similar
to the behavior of host-galaxy dust and the color parameters of other fitters.  The
$x(1)$ parameter affects the light-curve width and color around peak, similar to the stretch ($x_1$) and $\Delta$ parameters of SALT2 and MLCS.
The $x(2)$ parameter affects peak colors  in a fashion inconsistent with dust ($g^{0.25}-i^{0.25}$, $r^{0.25}-i^{0.25}$, $z^{0.25}-i^{0.25}$
are positively correlated), controls the near-UV light curve width, and influences the light-curve decline 20 to 30-days after peak brightness.
The $x(3)$ parameter most notably affects peak color and the light-curve shape through all phases of the $g^{0.25}$ band.
The K13 light curve parameters capture light-curve diversity distinct from those of SALT2;
Figure 10 shows  plots of SALT2 versus K13 light-curve parameters.
The absolute magnitude at peak $B$-band brightness is taken to be an unknown function of
a set of 15 light-curve parameters;
after modeling the function as a Gaussian process and training, the absolute magnitude can be determined to a dispersion as low as 0.09 mag.
The larger number of light-curve parameters (and their principal component compression)
that reduce the dispersion
may be sensitive to the pertinent information encoded in the host-galaxy parameters.

In this article we look for dependence of the K13 light-curve parameters and Hubble residuals
on host-galaxy parameters.  K13 use a supernova dataset analyzed in C13b, 
so the absence of correlations between Hubble residuals and host-galaxy parameters
could provide positive evidence for the improved performance of the K13 method
with respect to supernova environmental biases.

\section{Data Set}
\label{data:sec}
Our sample starts with the 119 SNe Ia from K13.  These objects have spectrophotometric data sets obtained by
the SNfactory with the SuperNova Integral Field
Spectrograph \citep[SNIFS,][]{2004SPIE.5249..146L}
that had been fully processed as of early 2011.
The instrument has a fully filled $6\farcs 4 \times 6\farcs 4$ spectroscopic field of view
subdivided into a grid of $15 \times 15$ spatial elements and a
dual-channel spectrograph covering 3200--5200~\AA\ and 5100--10000~\AA.
The median signal-to-noise within 2.5-days of peak brightness is 10.2 per 2.4 \AA\ bin.
The observing strategy provides observations of each supernova every 2 to 4 days up to $\sim 45$
days after maximum light.
The spatial information provided by SNIFS
enables photometric flux extraction for each spectral resolution element, making
the reduced data spectrophotometric
\citep{2011MNRAS.418..258B, 2013A&A...549A...8B}.  Flux-calibrated multi-band light curves are composed
of synthetic photometry calculated by integrating spectra
over the transmission function of observer optical $griz$ bands.

We are interested in the light-curve parameters and Hubble residuals of the supernovae analyzed in K13 (the
same studied by C13b).
The objective of K13 was to calibrate absolute magnitudes at peak brightness based on light-curve shapes and colors.
The goal of this article is to examine possible biases in Hubble residuals, not to minimize the dispersion as in K13;
we therefore use a different sample selection increasing the number of supernovae used in this analysis to improve the statistics.

The supernova subsamples considered in our analysis are summarized in Table~\ref{samples:tab}.
Our {\bf full} sample requires $>16$ points over all bands  with
$S/N>50$: this ensures temporal coverage to inform the
regression of the multi-band light curves from roughly ten days before to forty days after $B$ maximum light. 
A $z<0.1$ requirement reduces sensitivity of Hubble residuals on the cosmological
parameters $\Omega_M$ and $\Omega_\Lambda$.
In K13, SN \#13 was identified as having one spectrum produce synthetic photometry
discrepant with the rest of the light curve (see their Figure~3 for the effect).  
For consistency with the K13 we take the very conservative step of rejecting any supernova
with  at least
one synthetic photometry point $>5\sigma$ away from the  mean regressed light curve predicted by the Gaussian process,
even though such spectra already had been internally flagged as having processing errors.
Only six objects (each with one outlying point), SNe \#8, \#14, \#85, \#98, \#99, and \#111
are thus culled.

\begin{deluxetable}{clc}
\tablecolumns{3}
\tablewidth{0pc}
\tablecaption{Supernova Samples \label{samples:tab}} 
 \tablehead{
\colhead{Sample}&\colhead{Selection Criteria}&\colhead{Number}}
\startdata
Initial& SNfactory data fully processed by early 2011 & 119 \\ \hline
Full & $\ge 16$ photometric points with $S/N \ge 50$ &  \\
 & $0.015<z< 0.1$& \\
 & Removal of SN \#13 identified in K13 as having spurious photometry&\\
 & No photometry $>5\sigma$ from regressed mean &103 \\ \hline
Early & In Full sample &  \\
& Photometry at least 2 days before $B$ maximum & 64\\
\enddata
\end{deluxetable}

The above requirements leave a full sample of 103 supernovae.
To ensure that biases are not introduced through extrapolation in the Gaussian process regression, we define the {\bf early} sample as the subset of 64 supernovae from the full sample with data at least two days before maximum brightness.

In K13, the calibration of absolute magnitudes is trained for the specific restframe wavelengths probed by the
observation, and several filter sets were considered.  The dispersion in Hubble residuals of the validation sample of 43 SNe
was smallest for the
absolute-magnitude calibration for the $i$-band of the blueshifted Dark Energy Survey $griz$ filter set,
i.e.\ the rest-frame bands of a supernova at redshift $z=0.25$ covered by standard $griz$,
which amount to supernova-frame effective wavelengths of 378, 497, 602, and 708 nm.
These filters are annotated with a superscript $griz^{0.25}$.
As our objective is to see how well SN~Ia  distances could be calibrated without
restricting ourselves to a specific observer filter system,  
we use this particular training to look for the dependence of light-curve parameters
and Hubble residuals on host-galaxy properties.
The Hubble residuals are calculated as the difference between the true and
inferred peak magnitudes of the Gaussian-process model from Table~5 in K13;
the true magnitude being that regressed by the light-curve model, the inferred being that regressed from
light-curve shape and color parameters.
They perform four distinct analyses with different training and validation sets such that each supernova was
in a validation set  once
and only once.  For each supernova, the tabulated absolute magnitude is the one inferred while it was part
of the validation set; in this work the Hubble residuals of all supernovae are treated collectively even though
they were not analyzed with the same trained model.

The statistics of interest in this article cannot benefit and in fact suffer
from the overtraining of light-curve and absolute-magnitude models that can occur
in the training process of K13.
Not all supernovae were used in the trainings of the K13 model; ``training'' subsets were used to specify the supernova model, which
in turn were used to predict absolute magnitudes of  independent ``validation'' subsets.  
Each absolute magnitude used in the current article was determined while the supernova was in the validation set, i.e.\ the training did not include the supernova itself.  Any overtraining of the training set thus appears as an added source of error in the absolute magnitudes.
In the present article, the signals of interest are deviations of Hubble residuals that are correlated with host-galaxy properties;
no host-galaxy property information was used in the determination of the absolute magnitudes.

\section{Supernovae with Data Coverage Over Peak Brightness}
\label{analysis:sec}

The early sample is a fair sample with which to examine the K13 analysis and training:
supernovae lacking rise-time data
extends the Gaussian process regression outside of its expected range of applicability as
peak magnitudes would then be extrapolated rather than interpolated.
The light curves in extrapolated regions revert toward the mean function (the updated template of \citet{2007ApJ...663.1187H}
in the analysis of K13), and have uncertainties bounded by the Gaussian Process kernel model
rather than data. For this reason, the K13 training sample was restricted to supernovae with
data earlier than 2 days before maximum.

\begin{figure*}
   \centering
   \epsscale{2}
   \plotone{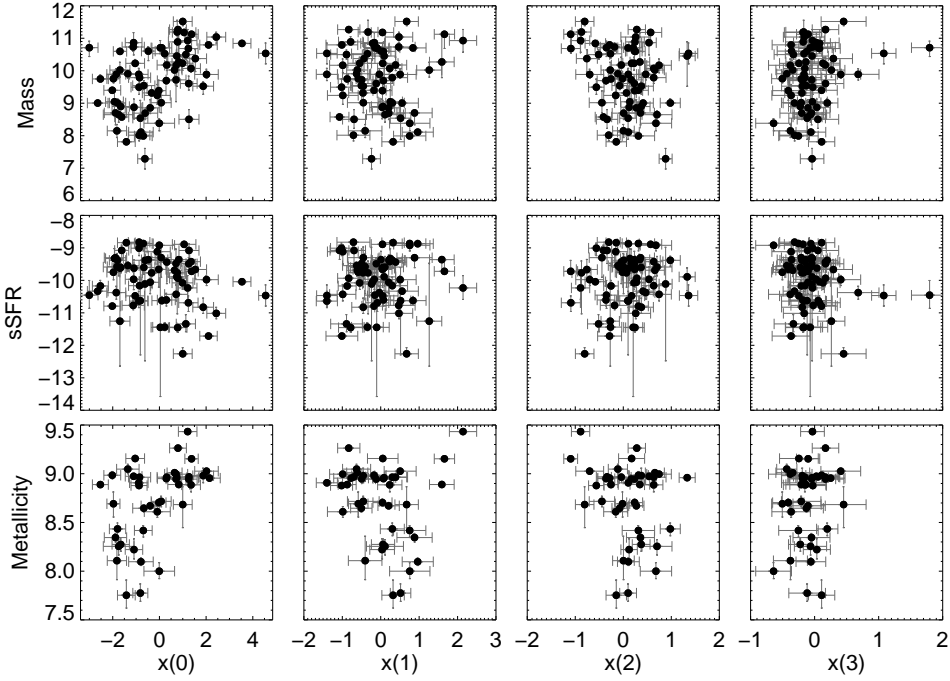} 
   \caption{
   Scatter plots of supernova light-curve parameters and host-galaxy mass $\log{(M/M_{\sun})}$, specific star formation rate  $\log{(sSFR)}$,
and metallicity $12+\log{(O/H)}$. \label{hosts:fig}}
\end{figure*}

We look for a correlation between  light-curve parameters and
the global host-galaxy properties of C13a in the early sample.
Note that each property has a different galaxy subset for which there is a measurement;
the metallicity measurement in particular requires emission lines and no
contamination from AGN emission.
Figure~\ref{hosts:fig} shows scatter plots from one of
the light-curve trainings.
The $x(0)$ light curve parameter is associated with color and exhibits
similar trends as the SALT2 $c$ parameter shown in C13b;
larger (redder) values trending
with higher mass and metallicity hosts as could be explained by host-galaxy dust absorption or a
color-metallicity dependence.
The $x(1)$ parameter is correlated with SALT2 $x_1$ and a smaller range of $x(1)$ values in low-mass and
low-metallicity hosts is apparent.
The $x(2)$ parameter has a larger color shift compared to $x(1)$,
broadens the post-maximum light curve, and shifts the height and phase of the
secondary maximum in $i^{0.25}$ and $z^{0.25}$.
There are no low  $x(2)$ values in low-metallicity hosts.
The $x(3)$ parameter is anti-correlated with $x_1$,  and is associated with variability in
UV light-curve brightness and shape
as well as the red secondary maximum.  The set of extreme positive $x(3)$ values are associated
with high-mass galaxies.

\begin{figure*}
   \centering
     \epsscale{1.4}
   \plotone{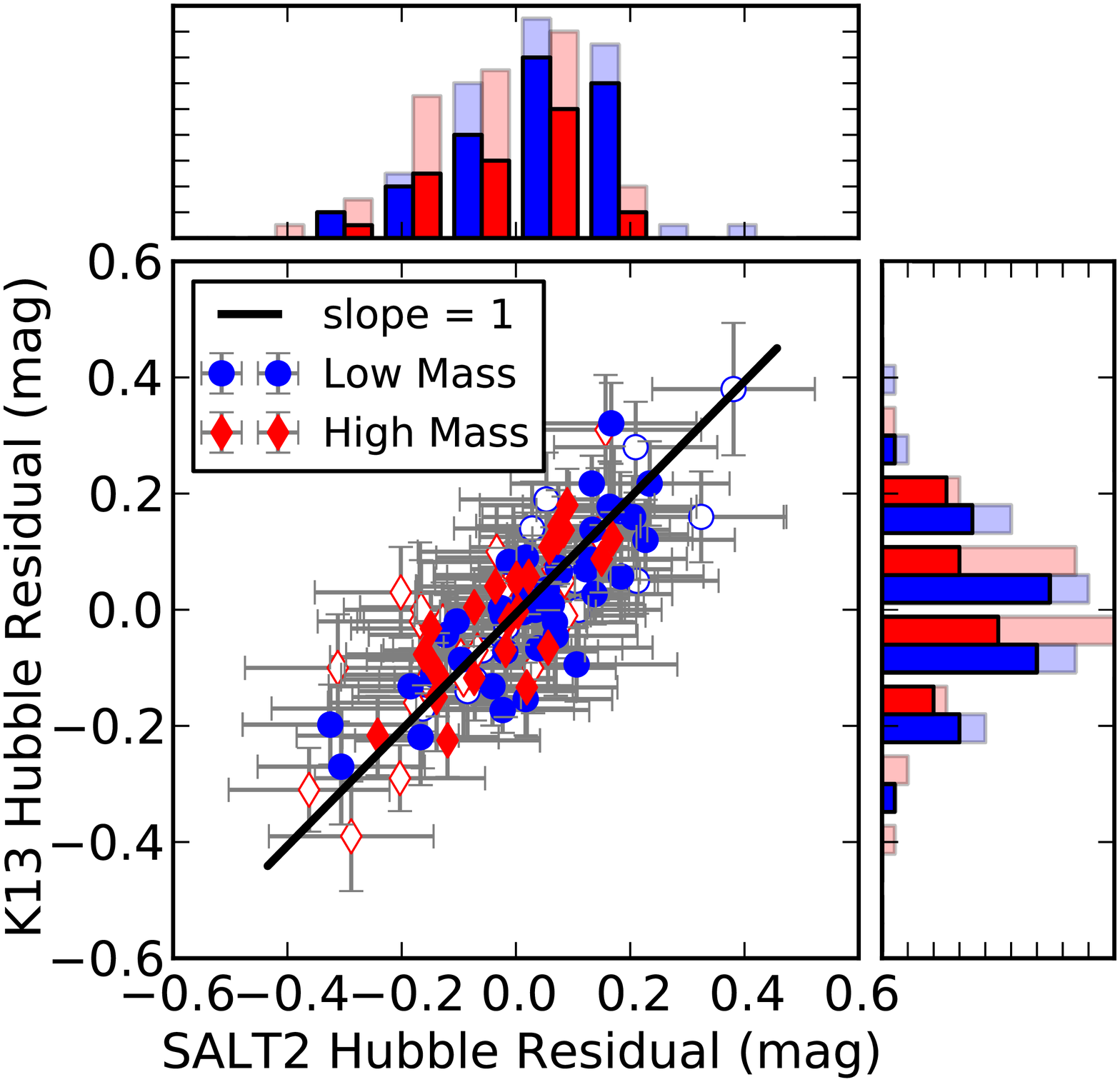}
   \caption{SALT2 versus K13 Hubble residuals
   for the full set of supernovae.  Subsamples from low-mass and high-mass hosts are shown as
    blue circles and red diamonds respectively.   The early sample is distinguished as solid and the
    complement as open.
   Histograms of the SALT2 and K13 Hubble residuals are shown in the top and right plots, with the early sample in solid.
\label{resres:fig}}
\end{figure*}

We use the results of C13
who apply a SALT2 fit and make magnitude corrections
on synthetic photometry using observer-frame non-overlapping boxcar filter functions whose
wavelength ranges correspond to Johnson-Cousins $B$, $V$, and $R$.
The K13 Hubble residuals are plotted against the SALT2 residuals
in Figure~\ref{resres:fig}, with the early sample distinguished with filled symbols.
Qualitatively, most of the solid points with SALT2 Hubble residual less than $-0.1$
have K13 residuals that are closer to zero (they lie above the $\mbox{slope}=1$ curve; also compare
the skewness from the negative sides of the histograms in Figure~\ref{resres:fig}.)
The positive correlation between the residuals shows that the two methods are subject to shared
``intrinsic dispersion'' that is irreducible for both.

Over the entire
early sample there is no significant bias between the supernova distances determined by K13 and SALT2:
the distribution of the difference between their Hubble residuals has mean $-0.004\pm0.010$ mag.  However, biases do appear when the early sample is divided
by global host-galaxy property.
For each host parameter, the population is split into two samples at boundaries established
by previous studies: at
 $\log{(M/M_{\sun})}=10.0$ for host-galaxy mass,
$\log{(sSFR)}=-10.3$ for specific star formation rate, and $12+\log{(O/H)}=8.8$ for metallicity. 
Not all hosts have measurements of these properties; for the early sample the number of supernovae
in the low- and high-bin subsets are $(37,24)$ for mass,
$(18,
43)$ for sSFR, and
$(18,
21)$ for metallicity.
For the full sample the numbers are
$(50,49)$ for mass,
$(38,
61)$ for sSFR, and
$(30,
32)$ for metallicity.

The  K13 minus SALT2 Hubble residuals in subsets defined by these splits are given
in Table~\ref{k13_salt_offsets:tab}.  The significance of the differences are at $1.6\sigma, 1.8\sigma$, and $2.7\sigma$ for mass, specific
star formation rate, and metallicity respectively.

\begin{deluxetable}{ccccccc}
\tablecolumns{7}
\tablewidth{0pc}
\tabletypesize{\tiny}
\tablecaption{Mean Difference Between K13 and SALT2 Hubble Residuals \label{k13_salt_offsets:tab}} 
 \tablehead{
\colhead{}&\multicolumn{2}{c}{ $\log{(M/M_{\sun})}$}&\multicolumn{2}{c}{$\log{(sSFR)}$}&\multicolumn{2}{c}{$12+\log{(O/H)}$}\\
\cline{2-3} \cline{4-5}  \cline{6-7}\\ 
\colhead{No Split}&\colhead{Low}&\colhead{High}&\colhead{Low}&\colhead{High}&\colhead{Low}&\colhead{High}\\
\colhead{(mag)}&\colhead{(mag)}&\colhead{(mag)}&\colhead{(mag)}&\colhead{(mag)}&\colhead{(mag)}&\colhead{(mag)}}
\startdata
\sidehead{Early Sample}
$-0.009 \pm 0.010$ &  $-0.020 \pm 0.013$ & $0.011 \pm 0.016$  &
$-0.025 \pm 0.015$ & $-0.000 \pm 0.010$  &
$-0.024 \pm 0.011$ & $0.024 \pm 0.015$  
\\
\hline
\sidehead{Full Sample}
$0.001 \pm 0.009$ &  $-0.021 \pm 0.011$ & $0.024 \pm 0.012$ & $0.006 \pm 0.014$ & $-0.003 \pm 0.011$ & $-0.007 \pm 0.010$ & $0.018 \pm 0.015$ \\
\enddata
\end{deluxetable}

The biases in K13 and SALT2 Hubble residuals imply that they should produce different Hubble residual steps.
The mean Hubble residual and intrinsic magnitude dispersion are fit for each subset defined by host-galaxy parameters
and any significant signal in the mean of the less-than sample minus the greater-than sample (i.e.\ the Hubble residual step)
is an indicator of population dependence.  A non-zero slope of a linear fit to the data also indicates dependence.

\begin{deluxetable}{ccccccc}
\tablecolumns{7}
\tablewidth{0pc}
\tabletypesize{\scriptsize}
\tablecaption{Hubble residual steps and slopes \label{offsets:tab}} 
 \tablehead{
\colhead{}&\multicolumn{2}{c}{ $\log{(M/M_{\sun})}$}&\multicolumn{2}{c}{$\log{(sSFR)}$}&\multicolumn{2}{c}{$12+\log{(O/H)}$}\\
\cline{2-3} \cline{4-5}  \cline{6-7}\\ 
\colhead{}&\colhead{offset}&\colhead{slope}&\colhead{offset}&\colhead{slope}&\colhead{offset}&\colhead{slope}\\
\colhead{Fit}&\colhead{(mag)}&\colhead{(mag dex$^{-1}$)} &\colhead{(mag)}&\colhead{(mag dex$^{-1}$)} &\colhead{(mag)}&\colhead{(mag dex$^{-1}$)} }
\startdata
\sidehead{Early}
$griz^{0.25}$ & $0.013 \pm 0.031$ & $-0.007 \pm 0.016$ & $-0.016 \pm 0.033$ & $0.044 \pm 0.022$ & $0.010 \pm 0.035$ & $-0.042 \pm 0.051$  \\
$gri^{0.25}$ & $0.020 \pm 0.031$ & $-0.023 \pm 0.016$ & $-0.036 \pm 0.031$ & $0.066 \pm 0.020$ & $-0.012 \pm 0.037$ & $-0.030 \pm 0.060$  \\
$riz^{0.25}$ & $0.020 \pm 0.037$ & $-0.034 \pm 0.018$ & $-0.049 \pm 0.036$ & $0.086 \pm 0.024$ & $-0.039 \pm 0.040$ & $0.213 \pm 0.079$ \\
SALT2 &  $0.047 \pm 0.039$ & $-0.047 \pm 0.013$ & $0.000 \pm 0.040$ & $0.030 \pm 0.015$ & $0.048 \pm 0.040$ & $-0.126 \pm 0.043$\\\hline
\sidehead{Full}
$griz^{0.25}$ & $0.045 \pm 0.026$ & $-0.028 \pm 0.013$ & $-0.037 \pm 0.027$ & $0.037 \pm 0.014$ & $0.029 \pm 0.030$ & $-0.068 \pm 0.040$ \\
SALT2 &  $0.095 \pm 0.030$ & $-0.047 \pm 0.013$ & $-0.048 \pm 0.031$ & $0.030 \pm 0.015$ & $0.059 \pm 0.032$ & $-0.126 \pm 0.043$  \\
\hline
\sidehead{C13b}
SALT2 &$   0.086 \pm    0.028$ &$  -0.043 \pm    0.014$ &$  -0.051 \pm    0.029$ &$   0.031 \pm    0.017$ &$   0.102 \pm    0.036$ &$  -0.106 \pm    0.046$\\
\enddata
\end{deluxetable}

\begin{figure*}
   \centering
     \epsscale{1.4}
   \plotone{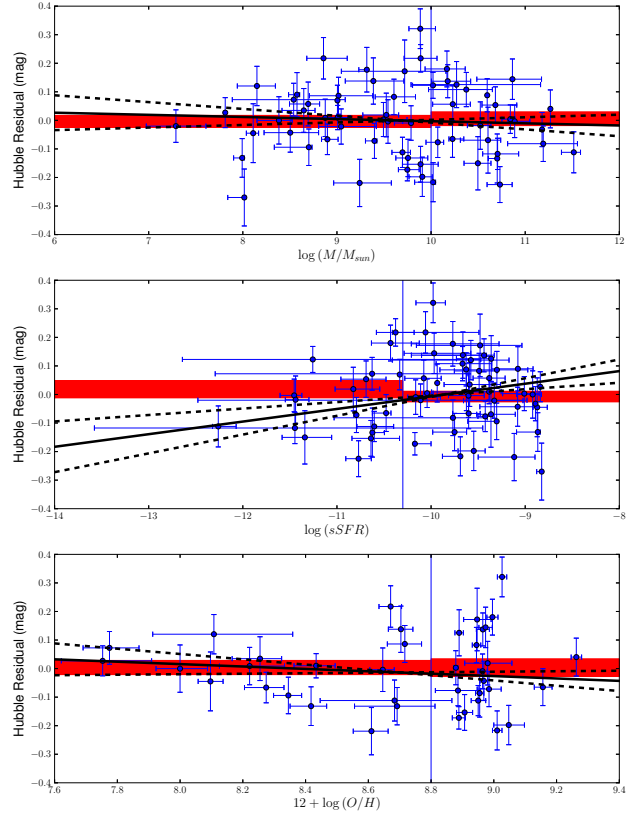}
   \caption{Hubble residuals of the $griz^{0.25}$-band analysis of K13 applied to the early sample,
   as a function of host-galaxy mass $\log{(M/M_{\sun})}$, specific star formation rate  $\log{(sSFR)}$,
and metallicity $12+\log{(O/H)}$.   Solid bands show the $\pm1\sigma$ measurement of the Hubble residual
weighted mean for each subset.  The solid line shows the best fit and the dashed lines the 1$\sigma$ range
of the linear model. \label{corr:fig}}
\end{figure*}

For the K13 4-band regressed approach, the Hubble residuals and slopes
are shown as a function of host galaxy mass, specific star formation rate,
and metallicity in Figure~\ref{corr:fig}.
The Hubble residual steps and slopes are listed in the Early/$griz^{0.25}$ row of Table~\ref{offsets:tab}.
The signals for step or slope have significance $<1\sigma$ for all cases except for
the slope in sSFR that has a $2\sigma$ significance.
We test for the presence of significant residual correlated structure
that may be not be captured in the simple step and slope statistics
by assuming an underlying covariance in the Hubble residuals described as $|a| k_{\nu = 3/2}(r;|l|)$
and fitting for $r$ and $l$, where
$a$ is a normalization parameter, $l$ the correlation length, $k_{\nu = 3/2}$ is the Mat\'ern function, and $r$
is the separation in the log galaxy parameter.  Probability distribution functions of the covariance parameters are built
from the best fits of multiple realizations of sets of  points generated based on the data and their uncertainties.  In all cases,
$a=0$ falls within less than 1 standard deviation of the mean; the data are consistent with having no correlation.

To examine the influence of wavelength coverage, the light-curve to absolute-magnitude relation is retrained as in K13
using the same filter set but omitting the extreme bands, i.e.\ for $gri^{0.25}$-data and
$riz^{0.25}$-data only.  The steps and slopes are listed as ``Early/$gri^{0.25}$'' and ``Early/$riz^{0.25}$'' in Table~\ref{offsets:tab}: no significant steps are
seen in either of these trainings but non-zero slopes (at $>3\sigma$) are found with sSFR for both.
For the $riz^{0.25}$ set in particular, the significance and size of the slope exceeds that of SALT2.

The SALT2 (v2.2.0) Hubble residual steps and slopes 
are listed as the ``Early/SALT2'' entry in Table~\ref{offsets:tab}.
The significance of the mass step is reduced to $1.4\sigma$.  Splits in the other
host-galaxy properties show similar weakening
of the significance of the step in the two samples.

Differences in the Hubble residual steps from different samples or light-curve analyses are listed for select pairs
in Table~\ref{offsets2:tab}.   
The steps of the various filter sets  are calculated using the same supernovae and data,
so their differences and uncertainties are not taken directly from Table~\ref{offsets:tab}
but are recalculated independently.
A relevant case is the effect of excluding the blueshifted $g$-band at 378 nm;
as noted above the $riz^{0.25}$ fit has a comparatively larger and significant slope with
sSFR than those of other runs.

\begin{deluxetable}{cccc}
\tablecolumns{4}
\tablewidth{0pc}
\tablecaption{Differences in K13 and SALT2 Hubble residual steps. \label{offsets2:tab}} 
 \tablehead{
\colhead{}&\multicolumn{3}{c}{Difference in Hubble Residual Step (mag)}\\
\colhead{Fits}&\colhead{$\log{(M/M_{\sun})}$}&\colhead{$\log{(sSFR)}$}&\colhead{$12+\log{(O/H)}$}
}
\startdata
\sidehead{Early Sample}
$griz^{0.25}-gri^{0.25}$ & $0.001 \pm 0.016$ &
$0.019 \pm 0.017$ &
$0.013 \pm 0.017$ 
\\
$griz^{0.25}-riz^{0.25}$ & $-0.002 \pm 0.020$ &
$0.028 \pm 0.018$ &
$0.041 \pm 0.018$ 
\\
$gri^{0.25}-riz^{0.25}$ & $-0.004 \pm 0.020$ &
$0.012 \pm 0.020$ &
$0.026 \pm 0.017$ 
\\
$griz^{0.25}$-SALT2 &$-0.032 \pm 0.019$ & $-0.025 \pm 0.019$ & $-0.048 \pm 0.021$\\
\sidehead{Full Sample}
$griz^{0.25}$-SALT2 &$-0.046 \pm 0.016$ & $0.015 \pm 0.017$ & $-0.023 \pm 0.018$ \\
\enddata
\end{deluxetable}

\begin{figure}
   \centering
    \epsscale{1}
   \plotone{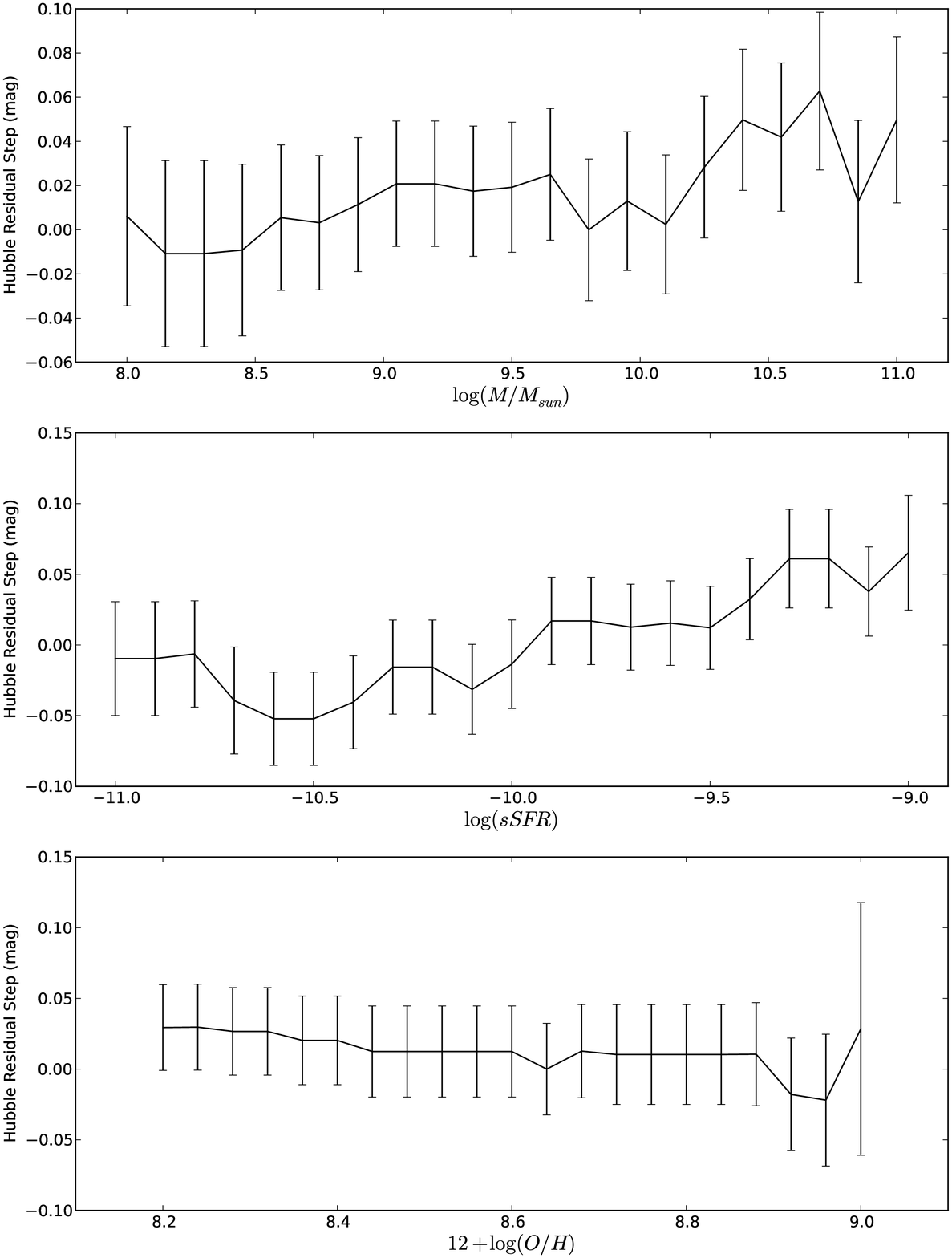}
   \caption{Hubble-residual steps  of the $griz^{0.25}$-band analysis of K13 applied to the early sample,
   for a range of parameter boundaries for  host-galaxy mass $\log{(M/M_{\sun})}$, specific star formation rate  $\log{(sSFR)}$,
and metallicity $12+\log{(O/H)}$.  \label{bound:fig}}
\end{figure}

The  host-galaxy-property  values used to separate the sample come from
published work that searched for systematic biases.  The objective of this article is to use K13 Hubble
residuals to probe
existing
positive detections of bias residuals, not to again mine for systematics. 
Nevertheless, we check whether
a different bias is evident in our sample.  Figure~\ref{bound:fig} shows  Hubble-residual steps for a range of parameter boundaries;
there are no $>2\sigma$
steps.

\section{Expanding the Supernova Sample}

Neither the K13- nor SALT2-determined distances of the early sample show significant $>3\sigma$
Hubble residual steps with respect to host-galaxy properties. 
We now consider how this subset behaves relative to larger subsets
with relaxed criteria for the first phase of SNIFS data.
This bridges the results of the early sample with C13b, who use the same data and SALT2 fits from the same parent supernova set to find a step at $>3\sigma$ with
host mass, consistent with previous studies \citep{2010ApJ...715..743K,2010MNRAS.406..782S,2010ApJ...722..566L,2011ApJ...740...92G, 2011arXiv1101.4269K,2011ApJ...743..172D}.
Our recalculation of SALT2 Hubble residual steps for the C13b sample are
given in Table~\ref{offsets:tab} as ``C13b/SALT2''; the results are consistent with C13b to 0.003 mag, the slight differences
being due to our fitting of separate (rather than joint) intrinsic dispersions for each population.

The principal feature that distinguishes the early sample is the requirement
for data at least $2$ days before peak. We here examine the effect of changing
the earliest phase required in the light curves.
While the K13 analysis is expected to be biased with underestimated uncertainties for supernovae without data coverage
at peak, this does provide
useful insight when
bridging SALT2 results of the early sample with that of the full sample.

\begin{figure*}
   \centering
 \epsscale{1.4}
    \plotone{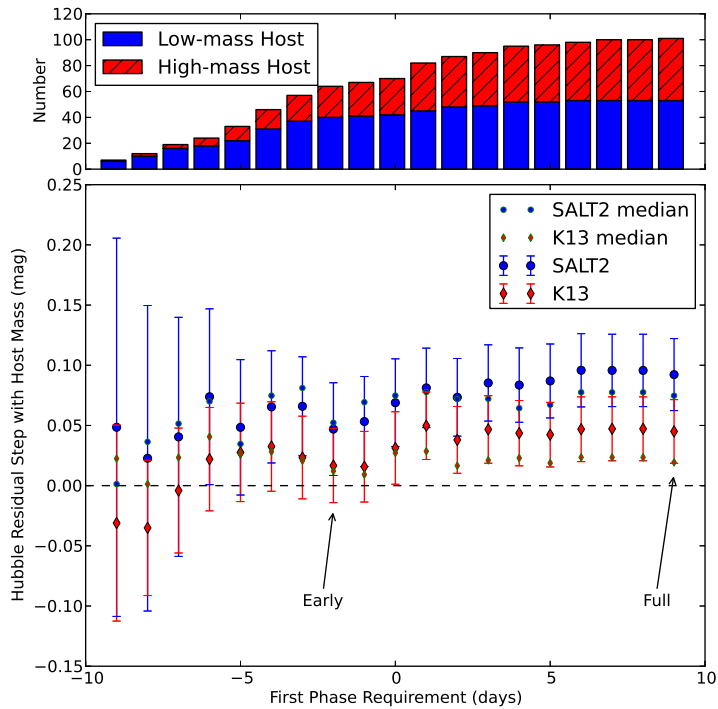}
   \caption{Hubble residual steps in the sample split by host-galaxy mass as a function of required phase of first observation.
   Each measurement includes the preceding sample of SNe.
   Blue circles are results for SALT2 and red diamonds
   for K13.  Larger symbols with error bars are steps calculated using the weighted-mean Hubble residuals, whereas the smaller symbols
   are median Hubble residuals.  The dashed line shows zero Hubble residual step between high- and low-mass hosts.
   The upper bar plot shows the number of supernovae from low-mass hosts (blue)
   and high-mass hosts (red hatched) included for each phase.  
   The samples representing the early and full samples are indicated with arrows.
\label{firstdate:fig}}
\end{figure*}

Figure~\ref{firstdate:fig} shows the Hubble residual steps with respect to host mass as a function of the required phase of first observation,
and the number of low-mass and high-mass that enter each calculation.
We first consider the asymptotic limit 
where the first phase must be $<9$ days after peak: this unconstraining requirement
accommodates all 103 supernovae in the full sample.  
Table~\ref{k13_salt_offsets:tab} shows that like the early sample, biases between K13 and SALT2 Hubble residuals
in the full sample only appear
after spliting by host-galaxy property.
K13 and SALT2 Hubble residual steps for the full sample are given in the ``Full'' rows of Table~\ref{offsets:tab}.
This larger set does have a
positive SALT2 mass-step detection at $3.2\sigma$;
the K13 result has a less significant step detection at $1.7\sigma$. The steps calculated
from median
Hubble residuals are 0.061 mag and $0.019$ mag for SALT2 and K13 respectively, within $1\sigma$ of the weighted-mean results.
Similar trends are seen with sSFR and metallicity. 
Note that the full sample, with a slightly smaller sample, gives results similar to C13b.

The results of the early sample are (marginally at $<2\sigma$)
 consistent with those of the full sample.  
The complement $(\mbox{full}-\mbox{early})$ sample has a step
of $0.181 \pm 0.052$, so the difference in the steps of the two samples is $0.127\pm 0.064$ mag.
The step of the early sample
is within $1.6\sigma$  of the full sample using SALT2.
The null Hubble-residual step evolves into a positive signal
with a slight increase in signal and slight decrease of noise. A bias in the SALT2 determination of corrected peak brightness between the early and full samples could be introduced
when including
or omitting rise-time data. 
The reapplication of SALT2 on artificial supernovae in the early sample created by removal of early data
shifts magnitudes by $-0.010 \pm 0.008$ mag and $-0.005 \pm 0.006$ for low- and high-mass hosts respectively
and a new mass step of  $0.040 \pm 0.042$; this does not
account for the 0.048 mag difference between the early and full samples.
Samples defined by a more stringent first-phase requirement than the early sample have a small number and a fraction lower than the asymptotic limit
of high-mass hosted supernovae. 
We have performed an extensive search for physical sources of bias between observing start and host-mass: none have been found. 
Statistics examined include the
first-phase versus local surface brightness, supernova magnitude and
color at first observation, the SALT2 $x_1$ light-curve shape parameter, and source of supernova discovery.

For K13, the step in the early sample
is within $1.2\sigma$ of that from  the full sample.
The K13 median steps are consistent with zero over all first-phase requirements.

The criterion for data at least $2$ days before peak for the early sample is based on the selection
of K13, fixed before any association with host-galaxy properties was made.  Figure~\ref{firstdate:fig} shows that relaxing
the criterion to include additional supernovae with coverage before peak does not affect the Hubble residual step.
Three supernovae with K13 Hubble residuals $<-0.29$ mag from high-mass hosts with first data phases between maximum and 3-days after maximum,
are
responsible for the disparity between mean and median and the increase in Hubble residual step for both K13 and SALT2. 

The differences between the K13 and SALT2 mass steps are calculated and
shown in Table~\ref{offsets2:tab} for the early and full samples.  The
correlation between K13 and SALT2 Hubble residuals discussed in
\S\ref{analysis:sec} must be accounted for: the quadratic sum of the two step uncertainties and the
simple
by-eye comparison of the error bars in  Figure~\ref{resres:fig} overestimate uncertainty.
The difference in the K13 and SALT2 mass steps for the early sample is at $1.7\sigma$ and for the late sample
$2.9\sigma$, suggesting that the lower mass step of K13 is significant.
The probabilities that the magnitude of the mass step is less for the K13 distances compared to SALT2 distances are
0.954 and
0.998 for the early and full samples respectively.

\section{SALT2 Hubble Residual Steps with K13 Parameters}
\begin{figure*}
   \centering
   \epsscale{2}
   \plottwo{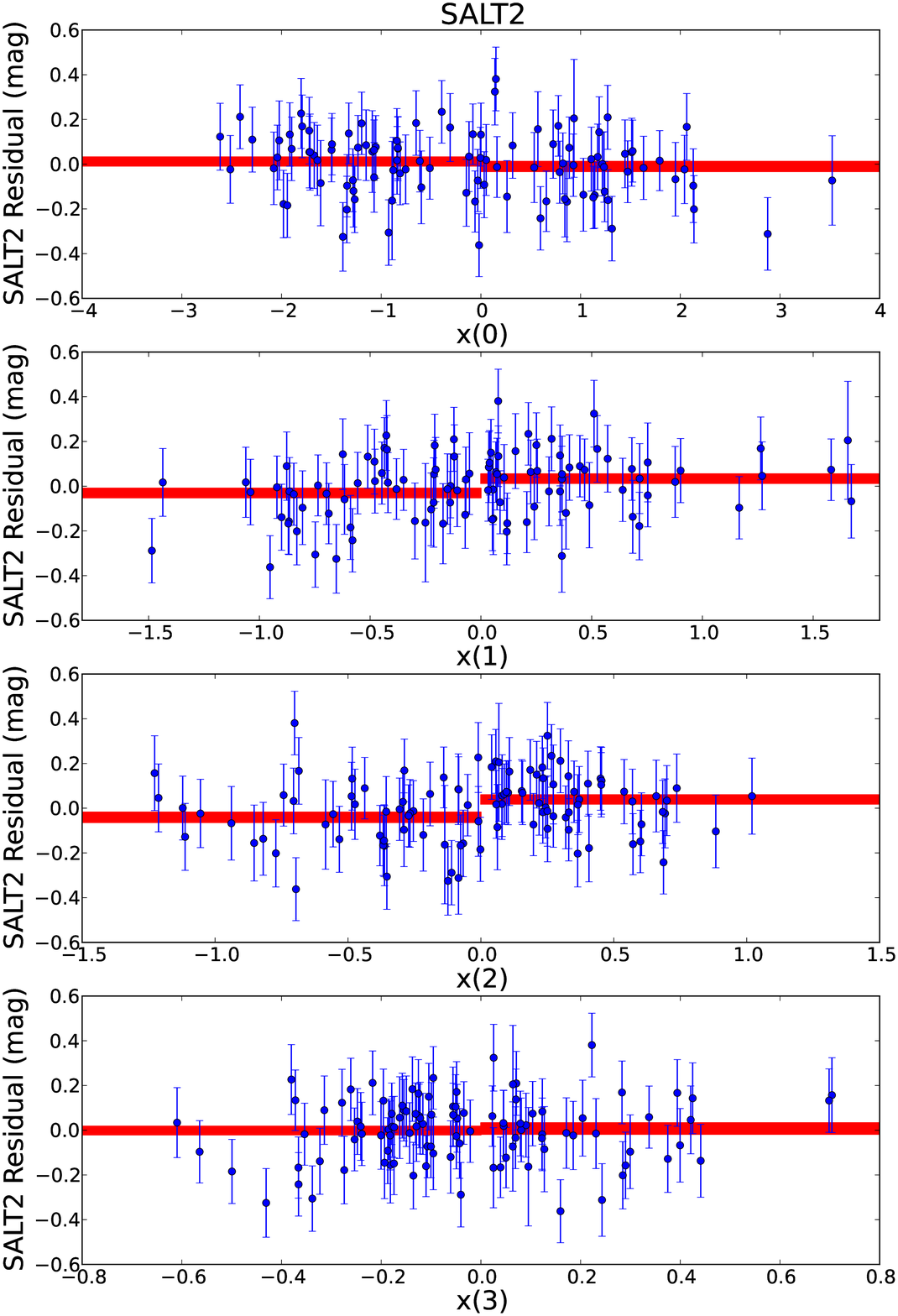}{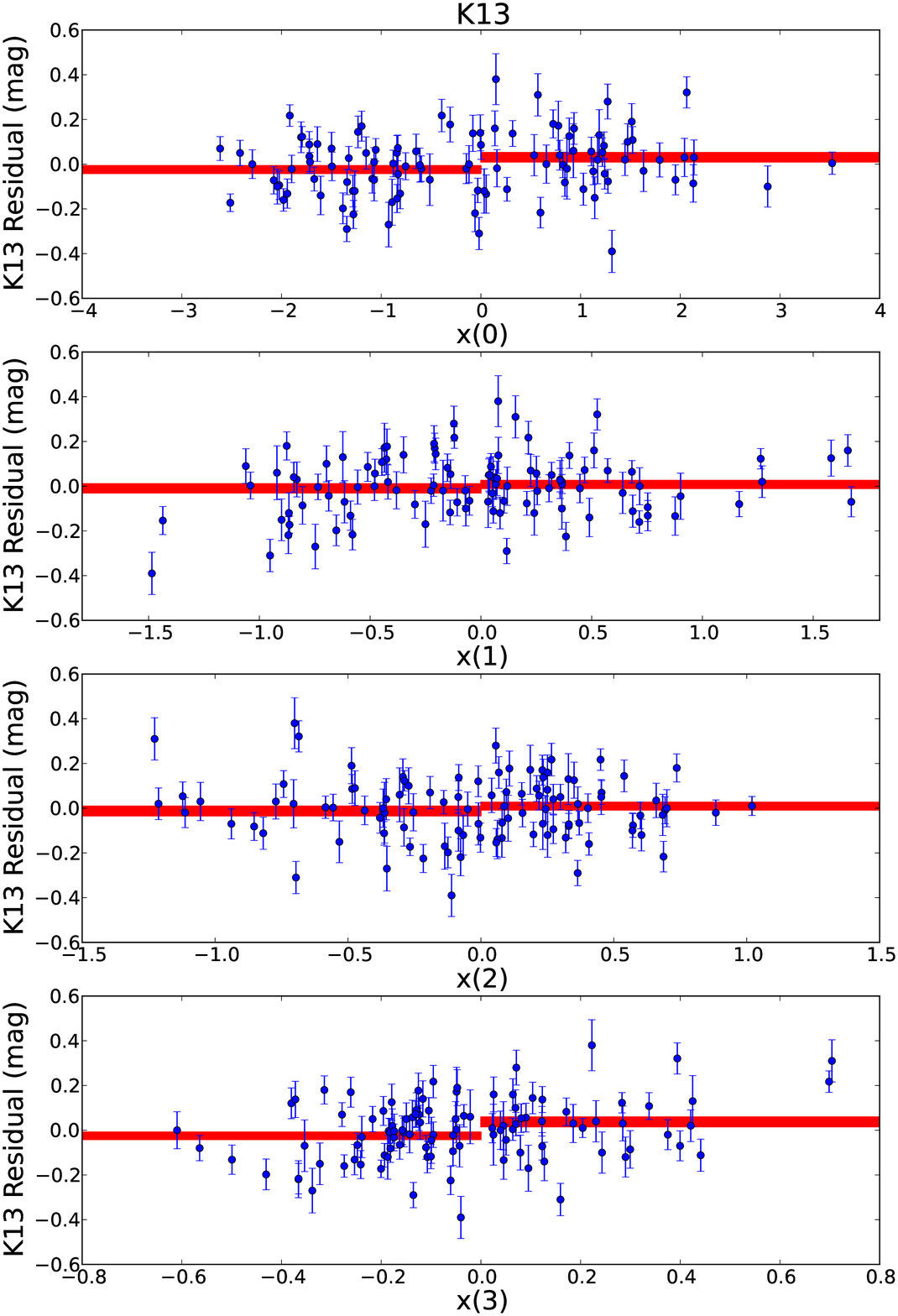}
   \caption{SALT2 (left) and K13 (right) Hubble residual steps in samples divided by the first four K13 parameters. Solid bands show the $\pm1\sigma$ measurement of the Hubble residual
weighted mean for each subset.\label{saltpca:fig}}
\end{figure*}

The differing SALT2 Hubble residual steps of the early and late samples may be more efficiently attributed to some underlying
light-curve parameter that is
not captured by SALT2 and that is not uniformly represented in the two samples; candidate parameters
include those identified in the Gaussian process analysis.
In a manner similar to the splitting of samples based on host-galaxy parameters,
we now calculate SALT2 Hubble residual steps dividing the full sample by each of the first four K13 parameters; results are given in Table~\ref{offsetsk13:tab}
for samples split by
 $x(0)$ through $x(3)$ in turn.
The Hubble residual steps are significant ($>2\sigma$)  in samples divided by  $x(1)$ and $x(2)$.
Hubble residuals as a function of these parameters are shown in Figure~\ref{saltpca:fig}, which also shows
as a point of reference and contrast K13 Hubble residuals as a function of these parameters.
The most significant step in K13 Hubble residuals for the same four parameters is $1.3\sigma$.
Recall that as noted in \S\ref{analysis:sec} and seen in Figures 6--9 of K13, the K13 parameters do capture diversity in the UV and
the secondary maximum at redder wavelengths that are not captured in the magnitude corrections using the SALT2 parameters $c$ and $x_1$.

\begin{deluxetable}{cccc}
\tablecolumns{4}
\tablewidth{0pc}
\tablecaption{SALT2 Hubble residual steps of the Full sample with K13 Parameters \label{offsetsk13:tab}} 
 \tablehead{
\colhead{ $x(0)$ step} & \colhead{$x(1)$ step} & \colhead{$x(2)$ step} & \colhead{$x(3)$ step}\\ 
\colhead{(mag)} & \colhead{(mag)} & \colhead{(mag)} & \colhead{(mag)}}
\startdata
$0.022 \pm 0.023$ &$-0.064 \pm 0.030$ &
$-0.080 \pm 0.030$ &
$-0.009 \pm 0.031$
\enddata
\end{deluxetable}

\section{Review and Discussion}
\label{discussion:sec}
Absolute magnitudes for a subset of Nearby Supernova Factory objects  determined using a novel regression method and
tabulated in K13 are used to measure the dependence of Hubble residuals on global host-galaxy properties.
Hubble residual steps with host mass, specific star formation rate, and metallicity are
 $0.013\pm 0.031$, $-0.016\pm0.033$, and
$0.010\pm0.035$ mag respectively: all consistent with zero step.
The SALT2 Hubble residual step for the same subset is  $0.047 \pm 0.039$ mag, consistent with a null step
at $1.2\sigma$.
As a point of comparison with a previous positive detection, \citet{2010ApJ...715..743K}
use an independent supernova sample and color--stretch magnitude
corrections to measure a mass
step of 0.11 mag at $2.3\sigma$.  This article's and their measurements are consistent at the $1.7 \sigma$ level using
K13, and at $1\sigma$ using SALT2 Hubble residuals.
For an expanded sample of supernovae with light curves unlike those in the K13 training,
the Hubble residual step with mass is
$0.045 \pm 0.026$ mag, which is
consistent with a null step at the 8\% level.  In contrast the color--stretch corrected magnitude step is
 $0.095 \pm 0.030$ mag, consistent with the null step at only 0.15\%.
The difference between the K13 and SALT2 mass steps for the full sample has an almost $3\sigma$ significance.

The analysis in this article differs with C13b in sample selection.
Expansion of the sample to include supernovae with later first data asymptotically
gives significant $(>3\sigma$) SALT2 Hubble residual steps with host-mass but not for K13 Hubble residuals.
To examine the difference between the early and full samples, we calculate the
difference between the SALT2 mass steps of the early and $(\mbox{full}-\mbox{early})$ samples to be 
$0.134\pm 0.065$, a difference this large has a 4.2\% chance of random occurrence.
The early and late samples do have different populations of objects, for example the early sample has a higher
ratio of low-mass hosts;  however, the difference in relative fraction of high- to low-mass hosts is not statistically significant.
We conclude that there is an evolution from null to positive-signal
SALT2 Hubble residual steps going from early to full samples, but that more statistics is required to determine whether
the samples are inherently drawn from different parent distributions.  K13 Hubble residual steps also evolve 
with sample, but asymptote
at $<2\sigma$ for the full sample.

The analysis in this article also differs with C13b in the inference of supernova absolute magnitude.
There is SALT2 Hubble residual dependence on both the K13 $x(1)$ and $x(2)$ parameters
at the $>2\sigma$ level though not meeting the $3\sigma$ standard for a positive detection)
that does not appear in K13 Hubble residuals.  We conclude that multi-band light curves do
convey information about supernova absolute magnitude that is not captured in SALT v.2.2.0.
The sample of supernova used to train SALT2  may have a different supernova population
than that of the SNfactory data, and its performance on the results of the ensemble average is expected to improve with
retraining using a subset of the SNfactory sample.
The GP analysis has an advantage in that it nulls out population effects by
having training and validation sets drawn from a common sample.
Nevertheless, if SALT2 did capture the diversity of light curves  no bias would arise due to differing training and validation populations
assuming that systematic changes in absolute magnitude are accompanied with changes in the time-evolving
SED.

The continued study of correlations between host-galaxy properties and Hubble
residuals is of importance even as tools that measure distances improve.  Correlations
serve as important diagnostics of potential biases related to progenitor population
that are not captured by light-curve parameters or absolute magnitude inference.
Better accuracy can be achieved from
local host-galaxy properties extracted from the region around the supernova
\citep{Rigault}, which can give a more accurate view of the progenitor environment
compared to global properties.

Physically, there is the expectation that multi-band light curves with broad wavelength coverage
transmit information about the progenitor system.
Due to limitations in current light-curve fitters, some of this information 
has only been noted indirectly through host-galaxy
properties.  This article shows that the novel light-curve analysis of K13 may be capturing
more aspects of SN~Ia diversity.  Application of the method to more supernovae will provide
better statistics to test this hypothesis.

\acknowledgements
We thank  Dan Birchall
for observing assistance, the technical
and scientific staffs of the Palomar Observatory, the High Performance
Wireless Radio Network (HPWREN), the National Energy Research Scientific
Computing Center (NERSC), and the University of Hawaii 2.2~m telescope.
We wish to recognize and acknowledge the significant cultural role and
reverence that the summit of Mauna Kea has always had within the
indigenous Hawaiian community.  We are most fortunate to have the
opportunity to conduct observations from this mountain. 
NC acknowledges support from the Lyon Institute of Origins under grant ANR-10-LABX-66.
This work was
supported by the Director, Office of Science, Office of High Energy
Physics, of the U.S. Department of Energy under Contract
No.~DE-AC02-05CH11231; by a grant from the Gordon \& Betty Moore
Foundation; in France by support from CNRS/IN2P3, CNRS/INSU, and PNC,
and by French state funds managed by the ANR within the Investissements d'Avenir program under reference ANR--11--IDEX--0004--02;
and in Germany by the DFG through TRR33 ``The Dark Universe.''  NERSC is
supported by the Director, Office of Science, Office of Advanced
Scientific Computing Research, of the U.S. Department of Energy under
Contract No.~DE-AC02-05CH11231.  HPWREN is funded by National Science
Foundation Grant Number ANI-0087344, and the University of California,
San Diego.

\end{document}